\documentclass[twocolumn,showpacs,amsmath,amssymb]{revtex4}
\usepackage{graphicx,url,hyperref,dcolumn,bm,lineno}
\hypersetup{colorlinks,citecolor=blue,linkcolor=red,urlcolor=blue}
\usepackage{graphicx}
\graphicspath{%
    {converted_graphics/}
    {/}
}
\begin{document}
\title{A statistical mechanics for immiscible and incompressible two-phase flow in porous media}

\author{Alex Hansen}
\email{alex.hansen@ntnu.no}
\affiliation{PoreLab, Department of Physics, Norwegian University of Science and Technology,
NTNU, N-7491 Trondheim, Norway}
\author{Eirik G.\ Flekk{\o}y} 
\email{flekkoy@fys.uio.no}
\author{Santanu Sinha}
\email{santanu.sinha@ntnu.no}
\affiliation{PoreLab, Department of Physics, University of Oslo, N-0316 Oslo, Norway}
\author{Per Arne Slotte}
\email {per.slotte@ntnu.no}
\affiliation{PoreLab, Department of Geoscience and Petroleum, Norwegian 
University of Science and Technology, NTNU, N-7491 Trondheim, Norway}
\date{\today {}}
\begin{abstract}
We construct a statistical mechanics for immiscible and incompressible two-phase flow in porous 
media under local steady-state conditions based on the Jaynes maximum entropy principle.
A cluster entropy is assigned to our lack of knowledge of, and control over, the
fluid and flow configurations in the pore space.
As a consequence, two new variables describing the flow emerge: The agiture, that 
describes the level of agitation of the two fluids, and the flow derivative which is conjugate to the saturation.
Agiture and flow derivative are the analogs of temperature and chemical potential in standard (thermal)
statistical mechanics. The associated thermodynamics-like formalism reveals a number of hitherto unknown relations between the variables 
that describe the flow, including fluctuations. The formalism opens for new approaches to characterize porous media with
respect to multi-phase flow for practical applications, replacing the simplistic relative permeability theory while still keeping the
number of variables tractable.
\end{abstract}
\maketitle

\section{Introduction}
\label{intro}

Flow in porous media\cite{b88,s11,b17,ffh22} is a large field that spans many 
disciplines, from biology and chemistry to soil science, geophysics, materials 
science. A central problem is to find a theoretical description of immiscible 
two-phase flow in porous media at scales large enough for the pores to be negligible in 
size, often referred to as the continuum or Darcy scale.  This is neither a new problem, nor are 
attempts at solutions new. The earliest attempt at solving the problem was that of Wyckoff and
Botset \cite{wb36} who regarded the flow of each of the immiscible fluids
as one moving in a pore space reduced by the other fluids, thus reducing its own ability to move.  
This approach, now known as relative permeability theory, is today the standard framework 
used for practical applications.

There has been no lack of attempts to go beyond this simple theory. Perhaps the
most advanced attempt to date is Thermodynamically Constrained Averaging Theory (TCAT) 
\cite{hg90,hg93,hg93b,nbh11,gm14}, based on thermodynamically consistent definitions 
made at the continuum scale based on volume averages of pore-scale thermodynamic quantities.
Closure relations are then formulated at the macro-scale along the lines of the 
homogenization approach of Whittaker \cite{w86}.  A key strength of TCAT is 
that all variables are defined in terms of pore-scale variables. A key disadvantage of TCAT 
is that many averaged variables are produced, and many complicated assumptions are 
needed to derive useful results. Another development based on non-equilibrium thermodynamics 
uses Euler homogeneity to define the up-scaled pressure.  From this, Kjelstrup et al.\ derive 
constitutive equations for the flow while keeping the number of variables down 
\cite{kbhhg19,kbhhg19b,bk22}. A challenge here is how to incorporate the structure
of the fluid clusters spanning many pores. 
There is also a class of theories based on detailed and specific assumptions 
concerning the physics involved. An example is Local Porosity Theory
\cite{hb00,h06a,h06b,h06c,hd10,dhh12}.  Another example is the Decomposition in
Prototype Flow (DeProf) theory which is a fluid mechanical model combined with 
non-equilibrium statistical mechanics based on a classification scheme of 
fluid configurations at the pore level \cite{vcp98,v12,v18}.

Recent work \cite{hsbkgv18,rsh20,rpsh22} explores a new approach to immiscible 
two-phase flow in porous media based on the Euler homogeneity theorem. It provides a 
transformation from the seepage velocity (i.e., average pore velocity) of
the more wetting fluid, $v_w$ and the less wetting fluid $v_n$ to
another pair of fluid velocities, the average seepage velocity of both fluids combined
$v_p$, and the {\it co-moving\/} velocity $v_m$.   The co-moving velocity 
appears as a result of the Euler homogeneity assumption.  
The transformation is reversible: knowing the average seepage velocity and
the co-moving velocity, one can determine the seepage velocity of each fluid,
$(v_p,v_m)\rightleftarrows (v_w,v_n)$.

The mapping from the average velocity and the co-moving velocity to the seepage
velocity of each fluid together with the assumption that the fluids are incompressible,
constitutes a closed set of equations when supplemented by constitutive equations for
the average velocity $v_p$ and the co-moving velocity $v_m$.  These two constitutive equations 
relate the two velocities to the driving forces, the pressure and saturation gradients.

The constitutive equation for the average flow velocity $v_p$ relates to recent findings
starting with Tallakstad et al.\ \cite{tkrlmtf09,tlkrfm09} who reported pressure drop proportional to 
the volumetric flow rate raised to a power $0.54\pm 0.08$ in a two-dimensional
model porous medium using a mixture of a compressible and an incompressible fluid
under steady-state conditions. Aursj{\o} et al. \cite{aetfhm14} found a power law with
a somewhat larger exponent using the same model porous medium, but with two incompressible 
fluids.  Similar results have since been observed by a number of groups, see
\cite{sh17,glbb20,zbglb21}.  There has also been a considerable effort to
understand these results theoretically and reproduce them numerically  
\cite{tkrlmtf09,tlkrfm09,gh11,sh12,shbk13,xw14,ydkst19,rsh19a,rsh19b,lhrt21,fsrh21}.   

The constitutive equation for the co-moving velocity $v_m$ has recently been studied by 
Roy et al.\ \cite{rpsh22} by reverse-engineering published relative permeability
data and by using a dynamic pore network model.  It turns out that this constitutive
equation has a surprisingly simple form.  We will return to this further on in this paper.   

We note that these two constitutive laws are based on the  {\it collective\/} behavior 
of both fluids combined.  It would seem to be an impossible task to disassemble these two
collective constitutive equations into one for each fluid.  However, this is precisely what the 
mapping $(v_p,v_m)\rightarrow (v_w,v_n)$ makes possible.  
  
The theory just described is modeled on thermodynamics as presented by Callen \cite{c74,c91}, 
which essentially consists of two ingredients: 1.\ an energy budget, 
i.e.\ the Gibbs relation and 2.\ an assumption that the variables describing the energy are 
Euler homogeneous functions.  In the theory of Hansen et al.\ \cite{hsbkgv18}, only the second
assumption is used.  Hence, the following question may be posed: Can one complete the analogy 
between immiscible fluid flow in porous media and thermodynamics? Finding a positive answer to
this question is the aim of this paper.   

The theory we are developing here assumes local steady-state flow.  This corresponds to local 
equilibrium in thermodynamics.  By steady-state flow we mean that the macroscopic averages 
characterizing the flow are well-defined.  This does not rule out that fluid clusters move, 
breaking up and merging at the pore level.    

Finding an analogy between thermodynamics and non-thermal macroscopic systems is not new.  
Edwards and Oakeshott in their ``Theory of Powders" \cite{eo89} had the same aim 
when setting up a theory for static granular media.  By making the assumption that
all packing configurations having the same packing fraction are equally probable,
thus constructing a microcanonical ensemble with the packing fraction playing the 
role of energy, an analogy with statistical mechanics was made.  As a result, a non-thermal
pseudo-thermodynamics would follow. The ensuing theory thus had the same structure 
as thermodynamics, but the variables have nothing to do with thermodynamics --- 
only the framework being the same.  

In order to arrive at an analogue to thermodynamics in the present case, 
we take as a starting point the information 
theoretical statistical mechanics of Jaynes \cite{j57}. This generalizes the principles of 
statistical mechanics from being a theory specifically for mechanical systems, e.g., 
assemblies of molecules, to a framework that can be implemented once certain criteria are in 
place, whatever the system. In essence the Jaynes approach generalizes the Laplace Principle of 
Insufficient Reason. Quoting Jaynes \cite{j57}, ``Laplace's `Principle of Insufficient Reason' 
was an attempt to supply a criterion of choice, in which one said that two events are to be 
assigned equal probabilities if there is no reason to think otherwise."  

Jaynes furthermore builds his approach on Shannon's interpretation of entropy as a quantitative 
measure of what one does {\it not\/} know about a system --- the less is known, the larger the entropy.  
From a set of properties that such a function of ignorance must have, Shannon managed to construct a 
unique one fulfilling them \cite{s48}.  One of these properties was that knowing nothing at all, 
this function would have its maximum when all possible states of the system would be 
equally probable, thus generalizing the Laplace principle of insufficient reason.  Jaynes then 
interprets our knowledge about the system as constraints on our ignorance, so that that the Shannon 
entropy should be maximized under these constraints.

Constructing a theory along these principles for immiscible two-phase flow in porous media leads to
a pseudo-thermodynamics describing the flow. This entails finding powerful relations between the 
variables describing the flow.  Furthermore, it introduces new variables.  One such new variable is
the {\it cluster entropy\/} associated with the shapes of the fluid clusters.  This allows us to define an
{\it agiture,\/} essentially measuring the level of agitation of the two fluids. We have chosen the 
name ``agiture" to emphasize that it is {\it not\/} a temperature. 
Another variable is the {\it flow derivative\/} which is the conjugate of the saturation. This variable is
an analogue of the chemical potential in ordinary thermodynamics. A third new variable is the {\it flow pressure\/}
which is conjugate to the porosity. 

Furthermore, statistical mechanics is more powerful than thermodynamics in that all of thermodynamics may be
derived from statistical mechanics but not vice versa.  This is also the case here.  For example, 
fluctuations in the macroscopic variables are accessible via statistical mechanics only.      

We review in the next section the Euler homogeneity approach of \cite{hsbkgv18}.  Here we describe the
system we consider, defining central concepts.  In particular, we derive the two-way mapping 
$(v_p,v_m)\rightleftarrows (v_w,v_n)$ and discuss the co-moving velocity $v_m$.  In Section \ref{statmech} we
construct the Jaynes statistical mechanics for immiscible flow in porous media, and from this a 
pseudo-thermodynamics.  In order to effectuate these ideas in practice, we need to define how we measure the 
relevant variables. This means defining averaging procedures, which should be done both in space and 
in time as already pointed out by McClure et al.\ \cite{mba21a,mba21b}. We go as far in this section as 
to define and exemplify the equivalents of the 
Maxwell relations in ordinary thermodynamics. The next section \ref{fluctuations} concerns saturation and 
porosity fluctuations. 
In Section \ref{pressure}, we discuss the relation between the agiture, flow derivative and the pressure gradient.
We do this by considering the internal balance in a porous medium having two regions with different matrix properties.     

We end the paper by a discussion and conclusion section. Here we list a number of questions that remain 
open.

\section{Review of Euler homogeneity approach}
\label{euler}

Before we turn to constructing the statistical mechanics, we review the
Euler homogeneity approach to immiscible and incompressible two-phase flow in porous media
first introduced by Hansen et al.\ \cite{hsbkgv18}.

\subsection{Representative Elementary Area}
\label{rea}

\begin{figure}[tbp] 
  \centering
  \includegraphics[bb=132 141 757 441,width=7cm,keepaspectratio]{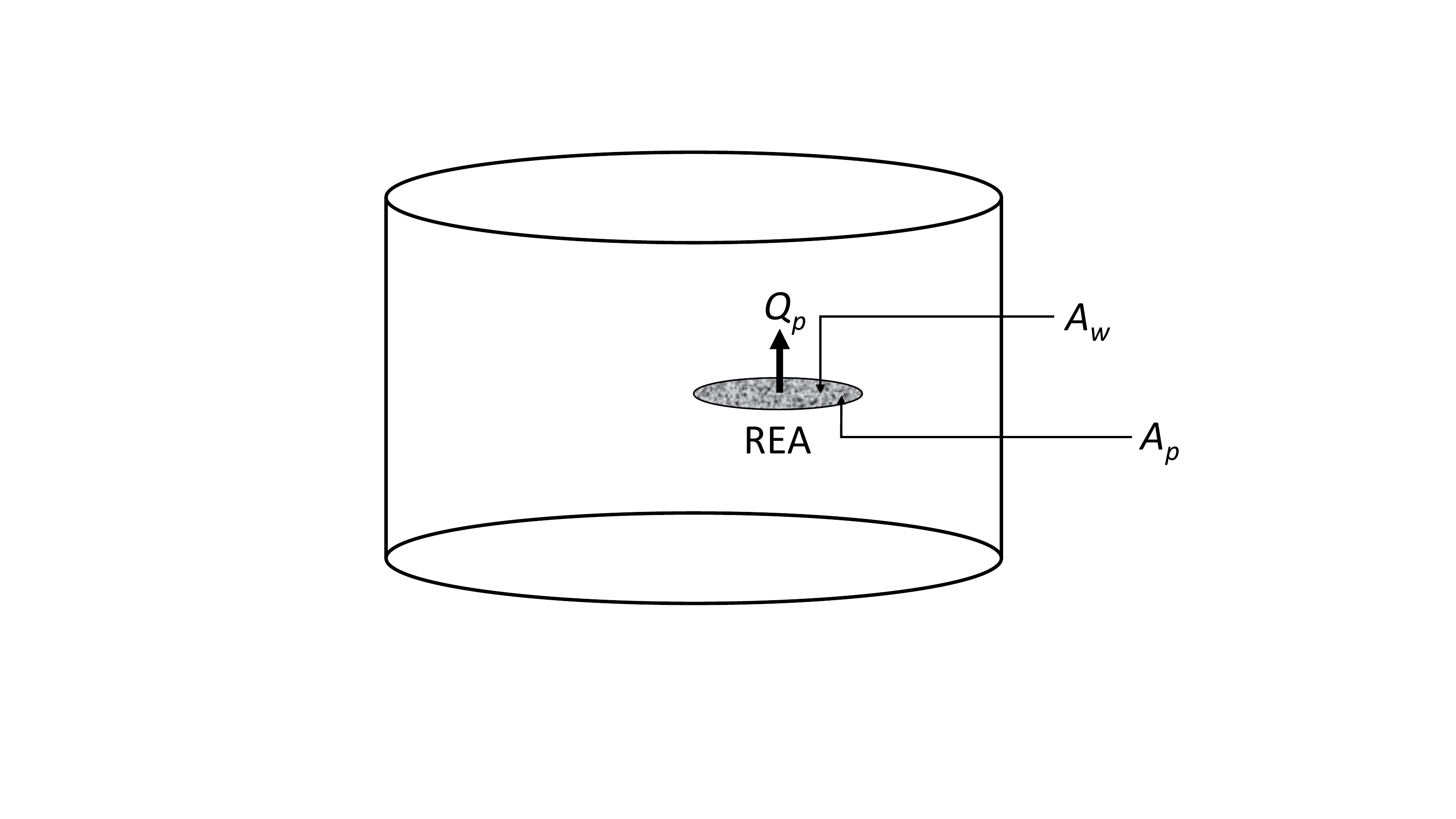}
  \caption{A Representative Elementary Area (REA) inside a porous medium plug.
There is a volumetric flow rate $Q_p$ passing though the REA which 
has a transverse pore area $A_p$, of which an area $A_w$ is covered by the 
wetting fluid.}
  \label{fig1}
\end{figure}

We imagine a porous medium plug as shown in Figure \ref{fig1}. It is homogeneous,
i.e., the local porosity and permeability fluctuate around well-defined averages.
The sides of the plug are sealed while the two ends remain open.  We ignore gravity.  
A mixture of two immiscible fluids are injected through the lower end, and fluids 
are drained at the upper end. Flow into the plug is constant, leading to steady-state
flow inside the plug \cite{esthfm13}.

Due to the sealed sides of the plug, the average flow direction is along the symmetry
axis of the plug.  We now imagine a plane cutting though the plug orthogonally to the 
average flow direction.  In this plane, we choose a point. Around this point, we choose 
an area $A$, e.g., bounded by a circle, as shown in Figure \ref{fig1}. We assume that the area 
is large enough for averages of variables characterizing the flow are well-defined, but 
not larger.  Furthermore, we assume the linear size of the area to be larger than
any relevant correlation length in the system.  This defines the {\it Representative 
Elementary Area\/} (REA) \cite{bb12} at the chosen point. We may do the same at 
any point in the plane, and we may do this at any other such plane.   
         
\subsection{Some definitions}
\label{definitions}

The REA has an area $A$. Part of this area is covered by the matrix material,
whereas another part is covered by the pores. This latter area, $A_p$, we will
refer to as the transverse pore area.  The porosity of the REA is then given by
\begin{equation}
\label{eq100}
\phi=\frac{A_p}{A}\;.
\end{equation}

There is a volumetric flow rate through the REA shown in Figure \ref{fig1}, 
$Q_p$. The seepage velocity of the fluids passing through the REA is then
\begin{equation}
\label{eq200}
v_p=\frac{Q_p}{A_p}\;.
\end{equation}

The flow consists of a mixture of two incompressible fluids, one being more wetting with 
respect to the matrix than the other one.  We will refer to this fluid
as the ``wetting fluid."  The less wetting fluid, we will refer to as the
``non-wetting fluid."  Each of them is associated with a volumetric
flow rate, $Q_w$ and $Q_n$, and we have 
\begin{equation}
\label{eq300}
Q_p=Q_w+Q_n\;.
\end{equation}
The transverse pore area of the REA may also be divide into an area associated with the
wetting fluid, $A_w$ and an area associated with the non-wetting fluid, $A_n$,
so that
\begin{equation}
\label{eq400}
A_p=A_w+A_n\;.
\end{equation}

We may define a seepage velocity for each of the two fluids,
\begin{eqnarray}
v_w&=&\frac{Q_w}{A_w}\;,\label{eq500}\\
v_n&=&\frac{Q_n}{A_n}\;.\label{eq600}\\
\nonumber
\end{eqnarray}
We define the saturations of the two fluids passing through the REA 
as
\begin{eqnarray}
S_w&=&\frac{A_w}{A_p}\;,\label{eq700}\\
S_n&=&\frac{A_n}{A_p}\;.\label{eq800}\\
\nonumber
\end{eqnarray}

We may now combine these equations (\ref{eq200}) ---(\ref{eq800}) to
expressing the average seepage velocity as 
\begin{equation}
\label{eq900}
v_p=S_w v_w+S_n v_n\;.
\end{equation}

We note that could have made these definitions based on the density of the two fluids,
$\rho_w$ and $\rho_n$ rather than volumes. 

The porosity $\phi$, saturations $S_w$ and $S_n$, and the seepage velocities $v_p$, $v_w$ and $v_n$
are variables that may be associated with the chosen point surrounded by the REA. Since 
there for every point in the porous medium one may associate an REA, these variables, which do not depend on
the size or shape of the REA, may be seen as continuous fields. 

The variables we have defined should be well-defined.  This is only possible if they are not fluctuating too strongly.  
Flooding processes typically generate fractal structures \cite{ffh22}. There are, however, length scales associated 
with the mechanisms controlling these structures \cite{mmhft21}, and beyond the largest of these length scales, 
they cease to be fractal. The same is true for steady-state flow which we consider here. We expect self averaging 
to take place beyond these length scales, i.e., the relative strength of the fluctuations shrinks with increasing 
size of the system.  Hence, we assume the plug and the REAs to be large enough for the fluctuations not to dominate.      

In the discussion that follows, we need to assign another property to the variables beyond being self averaging.  
We need them to be {\it state variables.\/} By this we mean that they depend on the flow properties there and then
and not the history of the flow.  Erpelding et al.\ \cite{esthfm13} studied this question experimentally and 
computationally, finding that this is indeed so.  

A last aspect to be considered is that of {\it hysteresis.\/} There may indeed be hysteresis 
in  the state variables.  Knudsen and Hansen \cite{kh06} studied immiscible two-phase flow under steady-state 
conditions using a dynamic pore network model.  They found that there are two transitions between 
two-phase flow and single-phase flow when the wetting saturation is the control parameter.  The transition 
between only the non-wetting fluid moving at low saturation to both fluids moving at higher saturation does 
not show any hysteresis with respect to which way one passes through the transition.  On the other hand, the 
transition between only the wetting fluid moving at high saturation and both fluids moving at lower saturation 
does show a strong hysteresis, see Figure 2 in Knudsen and Hansen \cite{kh06}.  This hysteresis is probably caused 
by this transition being equivalent to a first order or a spinodal phase transition.  It is well known
that such transitions in ordinary equilibrium thermodynamics lead to hysteretic behavior in
the state variables.  Hysteretic behavior signals that there are regions of parameter space
where the macroscopic state variables are multi-valued. Hence, the underlying microscopic physics 
has more than one locally stable mode.  

\subsection{Relations between the seepage velocities}
\label{seepage}

Hansen et al.\ \cite{hsbkgv18} made the weak assumption that the volumetric flow
rate through the REA is a homogeneous function of order one.  That is, if we scale  
$A\to \lambda A$ where $\lambda$ is a scale factor, the volumetric flow rate would scale 
in the same way, i.e.,  
\begin{equation}
\label{eq1000}
Q_p(\lambda A_w,\lambda A_n)=\lambda Q_p(A_w,A_n)\;.
\end{equation}
We have here assumed that $A_w$ and $A_n$ are {\it independent\/} variables. 
This means that that we may change the area 
$A$ of the REA by changing $A_w$, while keeping $A_n$ fixed or changing 
$A_n$ while keeping $A_w$ fixed.  This makes $A$ and $A_p$ defined in equation 
(\ref{eq400}) {\it dependent\/} variables. We refer to Hansen et al.\ \cite{hsbkgv18}
for details around this.  

By taking the derivative of equation (\ref{eq1000})
with respect to $\lambda$ and then setting $\lambda=1$, we find  
\begin{equation}
\label{eq1100}
Q_p(A_w,A_n)=\left(\frac{\partial Q_p}{\partial A_w}\right)_{A_n}A_w+
\left(\frac{\partial Q_p}{\partial A_n}\right)_{A_w}A_n\;.
\end{equation}
See Section 7.2 in Hansen et al.\ \cite{hsbkgv18} for a step-by-step demonstration 
of how these derivatives are done for a capillary fiber bundle model.

By invoking equations (\ref{eq200}), (\ref{eq700}) and (\ref{eq800}), we find
\begin{eqnarray}
\label{eq1200}
v_p&=&\frac{Q_p}{A_P}
= S_w\left(\frac{\partial Q_p}{\partial A_w}\right)_{A_n}+
S_n\left(\frac{\partial Q_p}{\partial A_n}\right)_{A_w}\nonumber\\
&=&S_w \hat{v}_w+S_n \hat{v}_n\;,
\end{eqnarray}
where we have defined
\begin{eqnarray}
\hat{v}_{w}=\left(\frac{\partial Q_p}{\partial A_{w}}\right)_{A_{n}}\;,\label{eq1300}\\ 
\hat{v}_{n}=\left(\frac{\partial Q_p}{\partial A_{n}}\right)_{A_{w}}\;.\label{eq1400}\\
\nonumber  
\end{eqnarray}
We will refer to $\hat{v}_w$ and $\hat{v}_n$ as the {\it thermodynamic velocities.\/}

Let us now switch to treating $S_w$ and $A_p$ as the independent variables. Simple
manipulations give
\begin{eqnarray}
\left(\frac{\partial}{\partial A_w}\right)_{A_n}
&=&\left(\frac{\partial }{\partial A_p}\right)_{S_w}
+\frac{S_n}{A_p}\ \left(\frac{\partial }{\partial S_w}\right)_{A_p}
\;,\label{eq1500}\\
\left(\frac{\partial}{\partial A_n}\right)_{A_w}
&=&\left(\frac{\partial }{\partial A_p}\right)_{S_w}
-\frac{S_w}{A_p}\ \left(\frac{\partial }{\partial S_w}\right)_{A_p}
\;.\label{eq1600}
\end{eqnarray}
Combining these two expressions with the definitions of the thermodynamic
velocities (\ref{eq1300}) and (\ref{eq1400}) gives
\begin{eqnarray}
\hat{v}_w=v_p+S_n\frac{dv_p}{dS_w}\;,\label{eq1700}\\
\hat{v}_n=v_p-S_w\frac{dv_p}{dS_w}\;,\label{eq1800}\\
\nonumber
\end{eqnarray}
where we note that $S_n$, ${\hat v}_w$, ${\hat v}_n$ and $v_p$ are 
all function of $S_w$ and not of $A_p$. 

We combine equations (\ref{eq900}) and (\ref{eq1200}),
\begin{equation}
\label{eq1900}
v_p=S_w v_w+S_n v_n =S_w \hat{v}_w+S_n\hat{v}_n\;.
\end{equation}
This does not imply that $v_w=\hat{v}_w$ and $\hat{v}_n=v_n$. 
Rather, the most general relation between them is
\begin{eqnarray}
\hat{v}_{w}=v_{w}+v_mS_{n}\;,\label{eq2000}\\
\hat{v}_{n}=v_{n}-v_mS_{w}\;,\label{eq2100}\\
\nonumber
\end{eqnarray}
where $v_m$ is the {\it co-moving velocity\/} 
\cite{hsbkgv18,rsh20,rpsh22}. 

Combining these two expressions with equations (\ref{eq1700})
and (\ref{eq1800}) expresses the {\it physical\/} seepage velocities $v_w$ and
$v_n$ in terms of the average seepage velocity $v_p$ and the 
co-moving velocity $v_m$,  
\begin{eqnarray}
v_w=v_p +S_n \left(\frac{dv_p}{dS_w}-v_m\right)\;,\label{eq2200}\\
v_n=v_p -S_w \left(\frac{dv_p}{dS_w}-v_m\right)\;.\label{eq2300}\\
\nonumber
\end{eqnarray}

We may invert equations (\ref{eq2200}) and (\ref{eq2300}), finding
\begin{eqnarray}
v_p=S_wv_w+S_nv_n\;,\label{eq2250}\\
v_m=S_w\frac{dv_w}{dS_w}+S_n\frac{dv_n}{dS_w}\;.\label{eq2350}\\
\nonumber
\end{eqnarray}

These four equations, (\ref{eq2200}) to (\ref{eq2350}), constitute a two-way
mapping $(v_p,v_m) \rightleftarrows (v_w,v_n)$.  This means that having
constitutive equations for $v_p$ and $v_m$, makes it possible to derive
constitutive equations for $v_w$ and $v_n$.  In other theories, such
as relative permeability theory, only
the mapping $(v_w,v_n)\to v_p$ is given, making it impossible to start from
a constitutive equation for $v_p$. 

How to measure these variables from experimental data, see Roy et al.\
\cite{rpsh22}, who studied the co-moving velocity in detail using
relative permeability data from the literature and from analyzing 
data obtained using a dynamic pore network model.  They found that the 
co-moving velocity takes the very simple form
\begin{equation}
\label{eq2400}
v_m=A+B\frac{dv_p}{dS_w}\;,
\end{equation}
where $A$ and $B$ are coefficients dependent on the capillary number. 
It is still an open question as to why it has this 
simple form.  We will in the following be able to give a partial answer.  

With these equations, the assumption that the two fluids are incompressible
and constitutive equations for $v_p$ and $v_m$, we have a closed set of equations
describing the flow. 

\section{Statistical Mechanics}
\label{statmech}
\subsection{Fluid configurations in a plug}
\label{entropy}

We show in Figure \ref{fig1} the porous plug that we discussed in Section \ref{rea}. We will
center our discussion on this plug in the following. 

\begin{figure}[tbp] 
  \centering
  \includegraphics[bb=132 141 757 441,width=7cm,keepaspectratio]{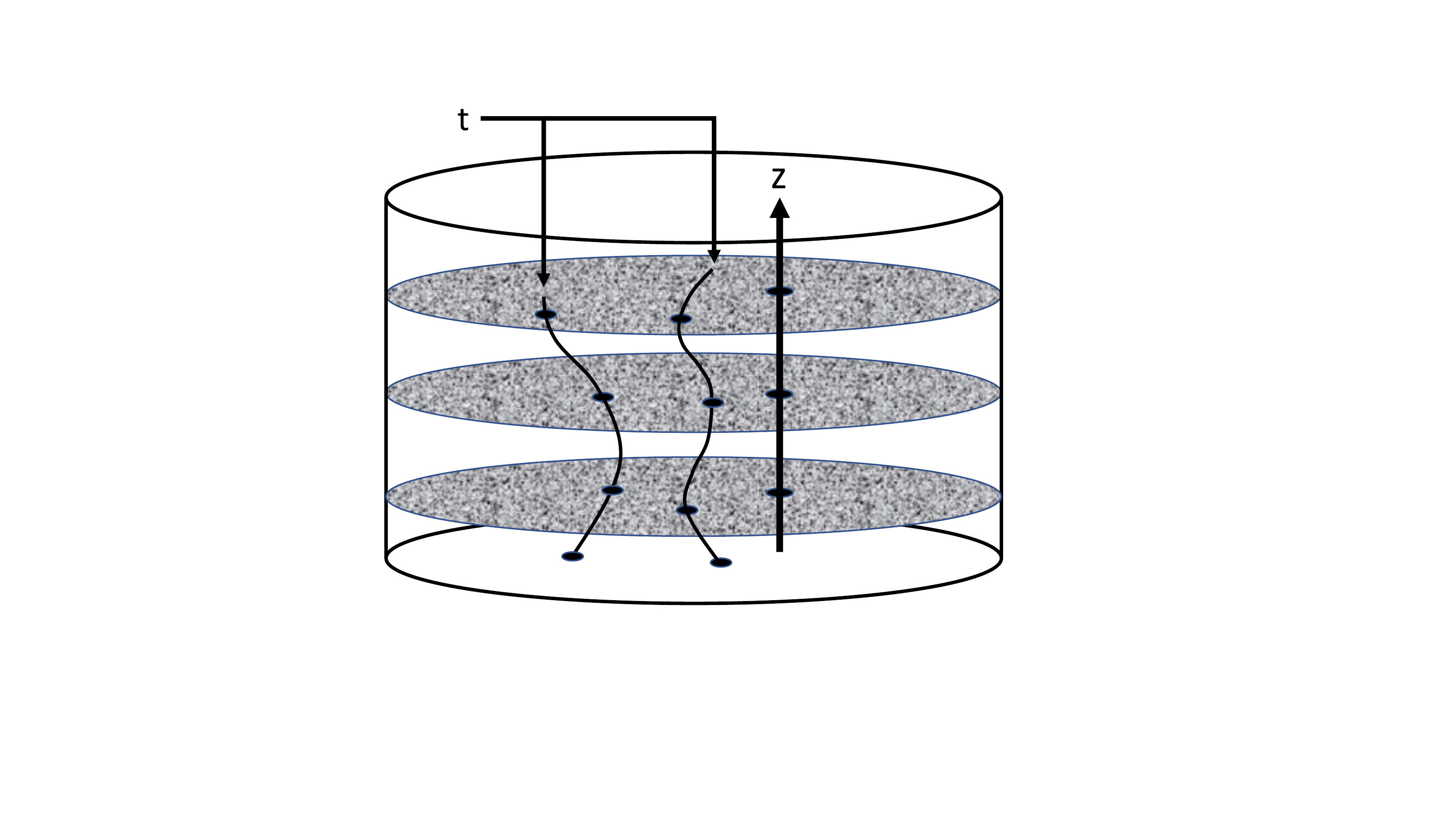}
  \caption{We show three planes orthogonal to the symmetry axis of the porous plug from 
Figure \ref{fig1} at different $z$ coordinates. We also show the trajectories of two
Lagrangian fluid elements. They started simultaneously at time $t=0$ at the plug inlet.
At time $t$ they are at the indicated positions.}
  \label{fig2}
\end{figure}

We assume that there is a volumetric flow rate ${\cal Q}_p$ passing through the plug under
steady-state conditions. 
This volumetric flow rate may be split into a wetting and a non-wetting volumetric
flow rate, ${\cal Q}_w$ and ${\cal Q}_n$ so that ${\cal Q}_p={\cal Q}_w+{\cal Q}_n$. We use
script characters to distinguish the flow through the entire plug from the flow through an REA. 
We show in Figure \ref{fig2} three planes orthogonal to the symmetry axis of the plug, 
i.e., orthogonal to the average flow direction. We introduce a coordinate $z$ along 
the symmetry axis of the plug measuring the distance from the plug's lower boundary
to any given plane. As the sides of the plug are sealed, we have that ${\cal Q}_p$ 
is the same through the three orthogonal planes --- or any other such orthogonal plane.  
The volumetric flow 
rate ${\cal Q}_p$ is a conserved quantity as we move along the $z$ axis. On the 
other hand, the volumetric flow rates of each of the two fluids, ${\cal Q}_w$ 
and ${\cal Q}_n$ are {\it not\/} conserved. Only their averages over several orthogonal planes
will remain constant. One may imagine the fluids being layered in the flow 
direction to realize this.  We also note that the transverse pore area $A_p$ will
fluctuate from plane to plane due to fluctuations in the local porosity. The transverse area 
associated with the wetting fluid, $A_w$, will also fluctuate for two reasons: $A_p$ fluctuates, but
more importantly because the fluid clusters fluctuate.    

We pick one of the planes, assuming it is at $z_0$. We assume $z_0$ is large enough so that there are 
no end effects in the plane from the inlet at $z=0$. One may at each point in the plane at $z_0$
assign a marker for whether the point is 1.\ in the solid matrix, 2.\ in the wetting fluid or 3.\ 
in the non-wetting fluid (thus ignoring the finite thickness of the interfaces and contact lines). 
We also assign a velocity vector to the point, which is zero if the point is in the solid 
matrix. This information collected for all points in the plane at $z_0$ at time $t$, we will 
refer to as the {\it fluid configuration\/} ${\cal X}={\cal X}(z_0,t)$ in this plane.    

We now imagine a {\it stack\/} of planes. We assume the neighboring planes are a distance $dz>0$ apart. 

In order to determine the flow configurations in each plane in the stack, it is not enough to know
it at entry plane of the stack, $z=z_0$. The reason for this is that we do not know the cluster 
structure inside the stack. However, the configurations in each plane are still measurable.  

If we move through the plug along the $z$-axis where $z_0 \le z \le z_1$, and the plug is long enough, we
will explore the space of possible configurations $\cal X$. We may do this at a fixed time $t$ (hence moving 
infinitely fast) or at a finite speed along the $z$-axis so that time is running.  We will in both cases explore 
the space of possible configurations.  If we choose a given plane and then average over the configurations
that pass it, we will not be averaging over the matrix structure, which will be fixed in the plane.  
However, if we imagine an {\it ensemble\/} of plugs each being a realization of the same statistical
pore structure, we will eventually see all configurations $\cal X$ also in this case.  

This leads us directly to defining a configurational probability density ${\tilde p}({\cal X})$.  We have also 
in the process sketched out an ergodicity assumption: the probability distribution for configurations in a 
given plane measured over an ensemble of plugs is the same as measuring it along different planes in a given plug.

Time seems to have fallen out of the description.  Time keeps track of the motion of each Lagrangian fluid element 
as we illustrate in Figure \ref{fig2}.  This is more information than we need.  All we need is to know the fluid
configurations in the planes.  For this purpose, {\it the $z$-axis acts as an effective time axis.\/} Hence, the reference
to an equivalent to the ergodic hypothesis which normally is a statement about time averaging vs.\ ensemble averaging.

Since we cannot reconstruct the configurations inside the stack given a knowledge of the configuration at the
entry plane at $z_0$, one may be inclined to reject the idea to interpret the $z$-axis as an effective time direction.
We point out that the same type of problem is encountered in relativistic mechanics of charged particles in strong fields 
where the position vs.\ time world lines of the particles are not single-valued functions of time \cite{f48,hr81}.  This 
means that it is not possible to reconstruct the later motion of such particles from knowing their configuration at
a given time.    

Statistical mechanics constitutes a calculational formalism based on the knowledge of the probability distribution
for configurations. The Jaynes maximum entropy principle provides a recipe for determining the configurational
probability density \cite{j57}.  Central to this principle is the definition of a function that quantitatively 
measures what we do {\it not\/} know about our system. This function is the Shannon entropy \cite{s48}. We
construct it for the present system as   
\begin{equation}
\label{eq2500}
\Sigma_{plane}=-\int d{\cal X} {\tilde p}({\cal X})\ln {\tilde p}({\cal X})\;,
\end{equation}
where the integral is over all hydrodynamically possible fluid configurations in the plane we focus on.  The 
task is to calculate this entropy and from it determine $\tilde p({\cal X})$.

We will refer to the entropy defined in equation (\ref{eq2500}) as the {\it cluster entropy.\/}  We have chosen the
name as it reflects the cluster structure that the fluids are making.  We emphasize that the cluster entropy is 
{\it not\/} the thermodynamic entropy defined in other work such as \cite{kbhhg19,kbhhg19b,bk22}.  Whereas there is
production of thermodynamic entropy in the plug as we are dealing with a driven system, there is no production of
cluster entropy as we are dealing with steady-state flow. 

The fluid states ${\cal X}$ are not discrete. Rather, they form a continuum.  Hence, we use an
integral in equation (\ref{eq2500}).  There are mathematical problems related to defining the
measure $d{\cal X}$ is this integral.  However, these difficulties we presume are of the same type encountered
in ordinary statistical mechanics.  We will not delve into these problems here, but rather 
just note their existence and that they have been solved in ordinary statistical mechanics.

\subsection{REA fluid configurations}
\label{REAentropy}

We have in Section \ref{rea} defined the REA. As for the entire plug, we may define 
a configuration $X$ for the REA as a hydrodynamically possible configuration within the
area $A$. The configuration in the plane where the REA sits is ${\cal X}$. Hence, $X$ is
a subset of $\cal X$. We also define the fluid configuration in the rest of the plane that
is not part of the REA, $X_r$.  We will refer to this part of the plane as the ``reservoir."
Hence, we have that 
\begin{equation}
\label{xxr}
{\cal X}=X\cup X_r\;.
\end{equation}
A central question is now, how independent are the configurations $X$ and $X_r$? If they are
independent, we may focus entirely on the REA configurations $X$ as we may write the 
configurational probability for the entire plane as
\begin{equation}
\label{tildeppp}
\tilde p({\cal X})=p(X)p_r(X_r)\;,
\end{equation}
where $p(X)$ is the configurational probability for the REA and $p_r(X_r)$ is the 
configurational probability for the reservoir.

 Fyhn et al.\ \cite{fsh22} have recently studied the validity of equation (\ref{tildeppp})
in a two-dimensional dynamic pore network model. By changing the size of the two-dimensional 
plug, while keeping the size of the REA fixed, they checked whether the statistical distributions 
of $Q_p$ and $A_w$ were dependent on the size of the plug.  Only a very weak dependency was found
which decrease with size.  It is therefore realistic to assume equation (\ref{tildeppp}) to be valid 
for large enough plugs and REAs.

\begin{figure}[tbp] 
  \centering
  \includegraphics[bb=132 141 757 441,width=7cm,keepaspectratio]{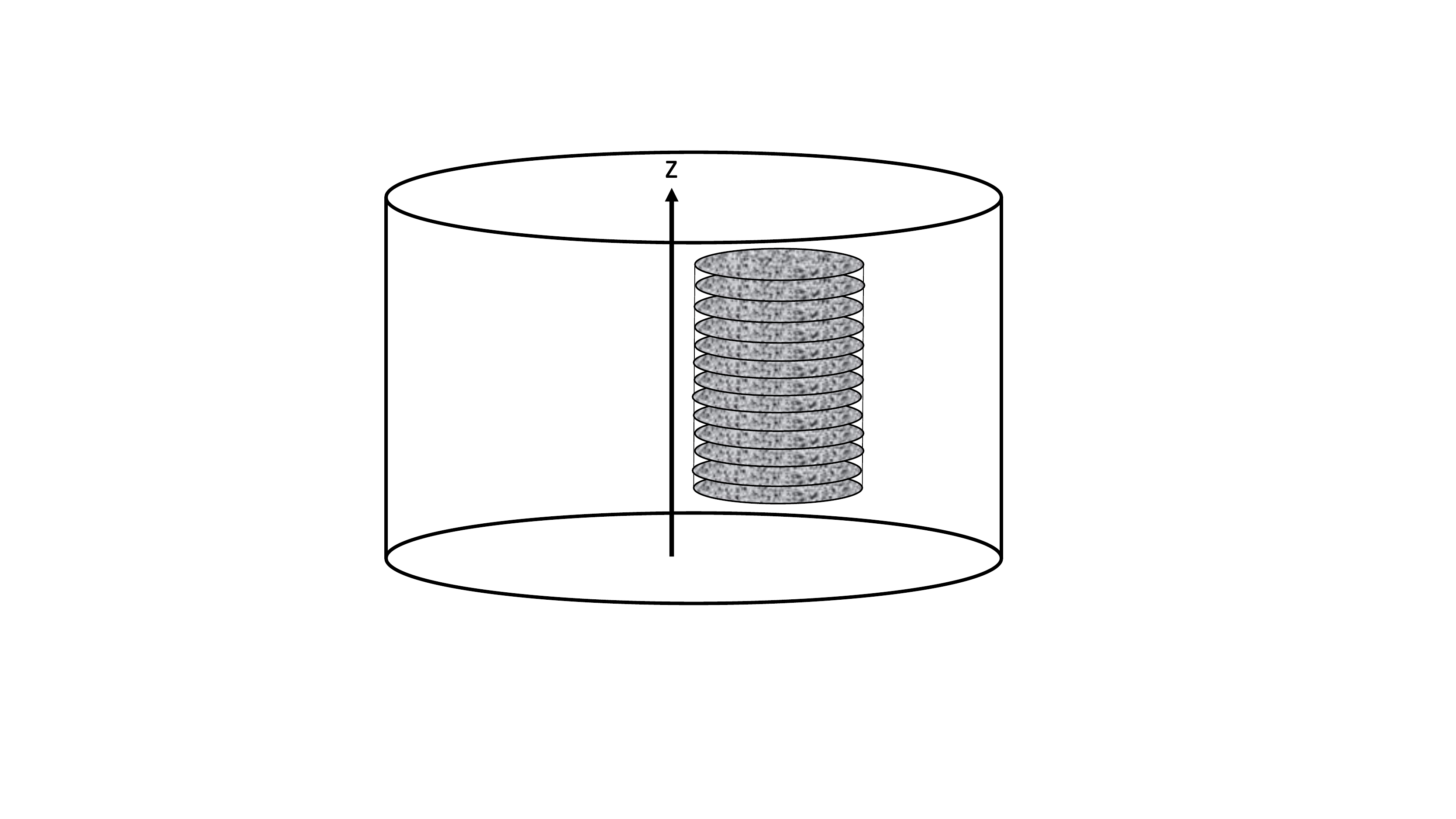}
  \caption{Averaging over a stack of REAs in the flow ($z$) direction. This ensures that 
the averaging is also over fluctuations in the pore structure.}
  \label{fig3}
\end{figure}

We proposed in Section \ref{entropy} to view the average flow direction through the plug, the $z$-axis
as a playing the role of a time axis.  Averaging over time thus corresponds to averaging over a stack of
REAs as shown in Figure \ref{fig3}.  When we in the next section refer to averaging, it is averaging over this stack we mean.

We may treat the interactions between the REA and the reservoir in different ways.  We may remove the REA stack
from the plug and treat it as a closed-off system. This amounts to treating the REA as a plug on its own.
We may leave it in the original plug, allowing it to freely interact with the reservoir.  We may allow the stack to
interact fully with the reservoir, but in such a way that the amount of wetting fluid is kept constant.  It is not possible 
to implement such constraints experimentally nor computationally, but theoretically it is. The same goes for the transverse
pore area: we may keep it constant theoretically, but not experimentally or computationally. 
Nevertheless, we will see examples of such ensembles in the following, as they correspond to different control parameters.   

\subsection{Jaynes maximum entropy approach}
\label{jaynes}

Following the Jaynes recipe, we need to maximize the cluster entropy in the REA, which is
\begin{equation}
\label{eq2550}
\Sigma=-\int dX p(X)\ln p(X)\;,
\end{equation}
under the constraints of what we know about the system.  The probability density $p(X)$ is defined in 
equation (\ref{tildeppp}).   

There are three variables that we measure, the volumetric flow rate through the REA, $Q_p$ and the transverse 
pore area covered by the wetting fluid, $A_w$ and the transverse pore area $A_p$. All three of them are averages 
over fluctuating quantities when performed as shown in Figure \ref{fig3}, i.e., we average over a stack of REAs. 

The average of the volumetric flow rate is               
\begin{equation}
\label{eq2600}
\int dX\ p(X) Q_p(X)=Q_p\;.
\end{equation}
where the $Q_p(X)$ is associated with fluid configuration $X$. Likewise,
the average wetting area is given by 
\begin{equation}
\label{eq2700}
\int dX\ p(X) A_w(X)=A_w\;,
\end{equation}
where $A_w(X)$ is the wetting area associated with configuration $X$. The third variable
we consider is the average transverse pore area  
\begin{equation}
\label{eq2750}
\int dX\ p(X) A_p(X)=A_p\;.
\end{equation}
All three variables $Q_p$, $A_w$ and $A_p$ are {\it extensive\/} in the area of the REA, $A$. The aim now is
to determine $p(X)$ based on the knowledge of these variable averages.  

Following Jaynes, we use the Lagrangian multiplier technique. We start by constructing the 
Lagrangian which we then will maximize, 
\begin{eqnarray}
\label{eq2800}
{\cal L}=&-&\int dX\ p(X) \ln p(X)\nonumber\\
&-&\Lambda\left(1-\int dX\ p(X)\right)\nonumber\\
&+&\lambda_Q\left(Q_p-\int dX\ p(X)Q_p(X)\right)\nonumber\\
&+&\lambda_w\left(A_w-\int dX\ p(X)A_w(X)\right)\nonumber\\
&+&\lambda_A\left(A_p-\int dX\ p(X)A_p(X)\right)\;,\nonumber\\
\end{eqnarray}
with the aim to determine $p(X)$. 

We assume that the three Lagrange multipliers $\lambda_Q$, $\lambda_w$ and $\lambda_A$ are 
{\it intensive\/} in the area of the REA, $A$.  The Lagrange
multiplier $\Lambda$, on the other hand, is extensive in the area $A$.  It follows that the cluster entropy 
$\Sigma$ is extensive in $A$.    

Necessary conditions for maximizing the Lagrangian (\ref{eq2800}) are 
\begin{eqnarray}
\frac{\partial {\cal L}}{\partial p(X)}&=&0\;,\label{eq2900}\\
\frac{\partial {\cal L}}{\partial \Lambda}&=&0\;,\label{eq3000}\\
\frac{\partial {\cal L}}{\partial \lambda_Q}&=&0\;,\label{eq3100}\\
\frac{\partial {\cal L}}{\partial \lambda_w}&=&0\;,\label{eq3200}\\
\frac{\partial {\cal L}}{\partial \lambda_A}&=&0\;.\label{eq3250}\\
\nonumber
\end{eqnarray}

One may ask why not additional information is included here, such as the average wetting flow rate, $Q_w$?
The answer to this lies in the Euler theory \cite{hsbkgv18}: there are three independent variables in
the problem for which we may fix their averages. We choose here $Q_p$, $A_w$ and $A_p$. Other averages
will then follow from the formalism we are about to develop.  

Equation (\ref{eq2900}) gives
\begin{equation}
\label{eq3300}
p(X)=e^{1-\Lambda}\ e^{-\lambda_Q Q_p(X)-\lambda_w A_w(X)-\lambda_AA_p(X)}\;,
\end{equation}
and the normalization condition, equation (\ref{eq3000}) gives
\begin{equation}
\label{eq3400}
e^{\Lambda-1}=\int dX e^{-\lambda_Q Q_p(X)-\lambda_w A_w(X)-\lambda_AA_p(X)} = Z\;,
\end{equation}
where we have defined the partition function $Z=Z(\lambda_Q,\lambda_w,\lambda_A)$. 

We are now in a position to calculate the cluster entropy combining equations (\ref{eq2550}) and (\ref{eq3300}),
\begin{eqnarray}
\label{eq3500}
\Sigma&=&-\int p(X) \ln p(X)\nonumber\\
&=&\ln Z(\lambda_Q,\lambda_w,\lambda_A)+\lambda_Q Q_p+\lambda_w A_w+\lambda_A A_p\;,\nonumber\\
\end{eqnarray}

We note that 
\begin{eqnarray}
Q_p(\lambda_Q,\lambda_w,\lambda_A)=-\left(\frac{\partial \ln Z }{\partial \lambda_Q}\right)_{\lambda_w,\lambda_A}\;,\label{eq3600}\\
A_w(\lambda_Q,\lambda_w,\lambda_A)=-\left(\frac{\partial \ln Z }{\partial \lambda_w}\right)_{\lambda_Q,\lambda_A}\;,\label{eq3700}\\
A_p(\lambda_Q,\lambda_w,\lambda_A)=-\left(\frac{\partial \ln Z }{\partial \lambda_A}\right)_{\lambda_Q,\lambda_w}\;.\label{eq3701}\\
\nonumber
\end{eqnarray}
These three equations may be solved to give $\lambda_Q$, $\lambda_w$ and $\lambda_A$ as functions of the three variables we know,
$Q_p$, $A_w$ and $A_p$. 
They also make equation (\ref{eq3500}) a triple Legendre transformation, changing the control variables 
$\lambda_Q \to Q_p$ and $\lambda_w \to A_w$ and $\lambda_A\to A_p$. Hence, the control variables of the flow
entropy are $\Sigma=\Sigma(Q_p,A_w,A_p)$:  
\begin{eqnarray}
\label{eq3501}
&&\Sigma(Q_p,A_w,A_p)=-\int p(X) \ln p(X)\nonumber\\
&=&\ln Z(\lambda_Q,\lambda_w,\lambda_A)
-\lambda_Q\left(\frac{\partial \ln Z }{\partial \lambda_Q}\right)_{\lambda_w,\lambda_A}\nonumber\\
&-&\lambda_W\left(\frac{\partial \ln Z }{\partial \lambda_w}\right)_{\lambda_Q,\lambda_A}
-\lambda_A\left(\frac{\partial \ln Z }{\partial \lambda_A}\right)_{\lambda_Q,\lambda_w}\;,
\end{eqnarray}

We define a new variable $Q_G=Q_G(\lambda_Q,\lambda_w,\lambda_A)$ through the equation
\begin{equation}
\label{eq3800}
Z(\lambda_Q,\lambda_w,\lambda_A)=e^{-\lambda_Q Q_G(\lambda_Q,\lambda_w,\lambda_A)}\;.
\end{equation}
It plays the role corresponding to that of a free energy in ordinary thermodynamics. 

Our next step is to invert equation (\ref{eq3500}) so that $Q_p$ becomes a function of $\Sigma$ rather 
the other way round. That is, we transform $\Sigma(Q_p,A_w,A_p)$ to $Q_p(\Sigma,A_w,A_p)$. 
Hence, we may write equation (\ref{eq3500}) or (\ref{eq3501}) as
\begin{eqnarray}
\label{eq3900}
Q_G(\lambda_Q,\lambda_w,\lambda_A)&=&Q_p(\Sigma,A_w,A_p)-\Sigma\frac{1}{\lambda_Q}\nonumber\\
&+&A_w\frac{\lambda_w}{\lambda_Q}+A_p\frac{\lambda_A}{\lambda_Q}\;,
\end{eqnarray}
where we have also used equation (\ref{eq3800}).  

We see that 
\begin{equation}
\label{eq4000}
\left(\frac{\partial Q_p(\Sigma,A_w,A_p)}{\partial \Sigma}\right)_{A_w,A_p}=\frac{1}{\lambda_Q}\;.
\end{equation}
Hence, we note that the following equation, which forms part of the right hand side of equation (\ref{eq3900}), 
constitutes a Legendre transformation, 
\begin{eqnarray}
\label{eq4100}
&&Q_F\left(\lambda_Q,A_w,A_p\right)=Q_p(\Sigma,A_w,A_p)-\Sigma\ \frac{1}{\lambda_Q}\nonumber\\
&=&Q_p(\Sigma,A_w,A_p)-\Sigma\ \left(\frac{\partial Q_p(\Sigma,A_w,A_p)}{\partial \Sigma}\right)_{A_w,A_p}\;.\nonumber\\
\end{eqnarray}
Here $Q_F$ corresponds to another free energy in ordinary thermodynamics.   

We rewrite equation (\ref{eq3900}) as
\begin{eqnarray}
\label{eq4750}
Q_G(\lambda_Q,\lambda_w,\lambda_A)&-&A_p\frac{\lambda_A}{\lambda_Q}\nonumber\\
&=&Q_F\left(\lambda_Q,A_w,A_P\right)+A_w\frac{\lambda_w}{\lambda_Q}\;.\nonumber\\
\end{eqnarray} 

The left hand side of this equation constitutes a Legendre transform,
\begin{equation}
\label{eq4780}
Q_M\left(\lambda_Q,\lambda_w,A_p\right)=Q_G(\lambda_Q,\lambda_w,\lambda_A)-\frac{A_p}{\lambda_Q}\ {\lambda_A}\;,
\end{equation}
as we have 
\begin{equation}
\label{eq4785}
\left(\frac{\partial Q_G(\lambda_Q,\lambda_w,\lambda_A)}{\partial \left(\frac{\lambda_A}{\lambda_Q}\right)}\right)_{\lambda_Q,\lambda_w}=A_p\;.
\end{equation}

Hence, we have now transformed equation (\ref{eq3900}) to
\begin{equation}
\label{eq4790}
Q_M\left(\lambda_Q,\lambda_w,A_p\right)=Q_F\left(\lambda_Q,A_w,A_p\right)
+A_w\frac{\lambda_w}{\lambda_Q}\;.
\end{equation} 

\subsection{Agiture, flow derivative and flow pressure}
\label{agiture}

Let us now define three new intensive variables built from the Lagrange multipliers 
$\lambda_Q$, $\lambda_w$ and $\lambda_A$,
\begin{eqnarray}
\theta&=&+\frac{1}{\lambda_Q}\;,\label{eq4150}\\
\pi&=&-\frac{\lambda_A}{\lambda_Q}\;,\label{4450}\\
\mu&=&-\frac{\lambda_w}{\lambda_Q}\;.\label{eq4400}\\
\nonumber
\end{eqnarray}

The first one, $\theta$, by its resemblance to temperature in ordinary statistical mechanics, 
we will name the {\it agiture.\/} The unit of the agiture is the same as volumetric flow rate. However, it is
an intensive variable.

The second one, $\pi$, we will name the {\it flow pressure.\/} This variable is the conjugate of the  
flow area $A_p$, 
\begin{eqnarray}
\label{eq4350}
\pi&=&\left(\frac{\partial Q_F(\theta,A_w,A_p)}{\partial A_p}\right)_{\theta,A_w}\nonumber\\
&=&\left(\frac{\partial Q_p(\Sigma,A_w,A_p)}{\partial A_p}\right)_{\Sigma,A_w}\;.
\end{eqnarray} 
The unit of $\pi$ is inverse velocity. Referring to equation (\ref{eq100}), we see that 
$A_p$ is a measure of the porosity $\phi$ and $\pi$ is therefore a velocity variable 
conjugate to the porosity. 

The third variable, equation (\ref{eq4400}) we name the {\it flow derivative.\/} We will explain
the name in the next section. As $\pi$, its unit is that of a velocity. It is the conjugate of the transversal 
wetting fluid area $A_w$,
\begin{eqnarray}
\label{eq4300}
\mu&=&\left(\frac{\partial Q_F(\theta,A_w,A_p)}{\partial A_w}\right)_{\theta,A_p}\nonumber\\
&=&\left(\frac{\partial Q_p(\Sigma,A_w,A_p)}{\partial A_w}\right)_{\Sigma,A_p}\;.
\end{eqnarray}
playing a role similar to a {\it chemical potential\/} in ordinary statistical mechanics. Through
equation (\ref{eq700}), we see that the flow derivative is the conjugate of the wetting 
saturation $S_w$. 

\subsection{Connection with Euler homogeneity approach}
\label{nonwetq}

We need to define one more variable,
\begin{equation}
\label{eq47920}
Q_N(\Sigma,\mu,A_p)=Q_M\left(\theta,\mu,A_p\right)-\Sigma(\theta,\mu,A_p)\theta\;,
\end{equation}
where
\begin{equation}
\label{eq47930}
\Sigma(\theta,\mu,A_p)=\left(\frac{\partial Q_M(\theta,\mu,A_p)}{\partial \theta}\right)_{\mu,A_p}\;.
\end{equation}
Combining this definition with equations (\ref{eq4100}) and (\ref{eq4790}) gives
\begin{equation}
\label{eq4791}
Q_N\left(\Sigma,\mu,A_p\right)=Q_p\left(\Sigma,A_w,A_p\right)-A_w\mu\;.
\end{equation}

We use the fact that both $Q_N(\Sigma,\mu,A_p)$ and $Q_p(\Sigma,A_w,A_p)$ are
homogeneous functions of order one in the extensive variables $\Sigma$, $A_w$ and $A_p$,
\begin{eqnarray}
Q_N\left(\lambda \Sigma,\mu,\lambda A_p\right)
&=&\lambda Q_N\left(\Sigma,\mu,A_p\right)\;,\label{eq4792}\\
Q_p\left(\lambda\Sigma,\lambda A_w,\lambda A_p\right)
&=&\lambda Q_p\left(\Sigma,A_w,A_p\right)\;.\label{eq4793}\\
\nonumber
\end{eqnarray}
We now set $\lambda=1/A_p$ and combine these two expressions with equation (\ref{eq4791}),
finding
\begin{equation}
\label{eq4794}
A_p Q_N\left(\sigma,\mu,1\right)=A_p Q_p\left(\sigma,S_w,1\right)-A_w\mu\;,
\end{equation}
where we also have used equation (\ref{eq700}) and we have defined the cluster entropy density
\begin{equation}
\label{eq47960}
\sigma=\frac{\Sigma}{A_p}\;.
\end{equation} 

We now divide equation (\ref{eq4794}) by $A_p$, noting that
\begin{eqnarray}
Q_N\left(\sigma,\mu,1\right)&=&v_N(\sigma,\mu)\;,\label{eq4795}\\
Q_p(\sigma,S_w,1)&=&v_p(\sigma,S_w)\;,\label{eq4796}\\
\nonumber
\end{eqnarray}
i.e., $v_N$ and $v_p$ are velocities. We recognize $v_p=Q_p/A_p$ from equation (\ref{eq200})
as the average seepage velocity of the two fluids.  Equation (\ref{eq4794}) then takes on the form
\begin{equation}
\label{eq4797}
v_N(\sigma,\mu)=v_p(\sigma,S_w)-S_w\mu\;.
\end{equation}
Let us now go back to equation (\ref{eq4300}) and use scaling relation (\ref{eq4793}) to find
\begin{equation}
\label{eq4798}
\left(\frac{\partial v_p(\sigma,S_w)}{\partial S_w}\right)_{\sigma}=\mu\;.
\end{equation}
This expression is the reason why we name $\mu$ the flow derivative.
We now compare these two equations (\ref{eq4797}) and (\ref{eq4798}) to
equation (\ref{eq2100}), which we reproduce here:
\begin{equation}
\nonumber
{\hat v}_n=v_p-S_w\left(\frac{d v_p}{d S_w}\right)\;.
\end{equation}
This equation is one of the central results derived by Hansen et al.\ \cite{hsbkgv18}. 
By comparison, we identify
\begin{equation}
\label{eq4801}
v_N={\hat v}_n\;,
\end{equation}
where the thermodynamic velocity of the non-wetting fluid, ${\hat v}_n$, is defined in
equation (\ref{eq1400}). Equation (\ref{eq4797}) may be written
\begin{equation}
\label{eq4799}
{\hat v}_n(\sigma,\mu)=v_p(\sigma,S_w(\sigma,\mu))-S_w(\sigma,\mu)\mu\;,
\end{equation}
and we have that 
\begin{equation}
\label{eq5100}
S_w(\sigma,\mu)=-\left(\frac{\partial {\hat v}_n(\sigma,\mu)}{\partial \mu}\right)_{\sigma}\;.
\end{equation}
These two equations are our central result.  It demonstrates that the {\it thermodynamic
velocity of the non-wetting fluid is the Legendre transform of the average seepage
velocity with respect to the wetting saturation and that the wetting saturation is minus the
derivative of the non-wetting thermodynamic velocity with respect to the flow derivative.\/}  
Equation (\ref{eq2100}) was derived based on the volumetric flow rate being a homogeneous function of the first kind
in the transverse pore area.  Now, we see this equation as a fundamental equation resulting 
from an underlying statistical mechanics, with variables $(\theta,\Sigma)$ and $(\mu,S_w)$,
and $(\pi,\phi)$ forming conjugate pairs.  

\subsection{Partition functions}
\label{partition}

The partition function is
given by
\begin{equation}
\label{eq3450}
Z(\theta,\mu,\pi)=\int\ dX\ e^{-Q_p(X)/\theta+\mu A_w(X)/\theta+\pi A_p(X)/\theta}\;.
\end{equation}
We may split the integration over states $X$ into an integral over $A_p$ and then over all
states $X$ that has a given $A_p$,
\begin{eqnarray}
\label{eq3451}
Z(\theta,\mu,\pi)&=&e^{-Q_G(\theta,\mu,\pi)/\theta}\nonumber\\
&=&\int_0^A\ dA_p e^{\pi A_p/\theta}\ Z(\theta,\mu,A_p)\;,
\end{eqnarray}
where we have defined 
\begin{eqnarray}
\label{eq3452}
&&Z(\theta,\mu,A_p)\nonumber\\
&=&\int\ dX\ \delta(A_p(X)-A_p)\ e^{-Q_p(X)/\theta+\mu A_w(X)/\theta}\nonumber\\
&=&\frac{1}{A}\int\ dX\ \delta(\phi(X)-\phi)\ e^{-Q_p(X)/\theta+\mu A_w(X)/\theta}\;,\nonumber\\
\end{eqnarray}
and where $\delta(\cdots)$ is the Dirac delta-function. We have used that $A\phi=A_p$,
see equation (\ref{eq100}). We have that
\begin{equation}
\label{eq3453}
Z(\theta,\mu,A_p)=\frac{1}{A}\ e^{-Q_M(\theta,\mu,A_p)/\theta}\;,
\end{equation}
where $Q_M(\theta,\mu,A_p)$ is defined in equation (\ref{eq4791}).  

We may now write partition function $Z(\theta,\mu,A_p)$ in equation (\ref{eq3452})
as
\begin{equation}
\label{eq3457}
Z(\theta,\mu,A_p)=\int_0^{A_p}\ dA_w\ e^{\mu A_w/\theta} Z(\theta,A_w,A_p)\;,
\end{equation}
where we have defined
\begin{eqnarray}
\label{eq3458}
&&Z(\theta,A_w,A_p)\nonumber\\
&=&\int\ dX\ \delta(A_w(X)-A_w)\ \delta(A_p(X)-A_p)e^{-Q_p(X)/\theta}\;,\nonumber\\
&=& \int\ \frac{dX}{A_p\ A}\ \delta(S_w(X)-S_w)\ \delta(\phi(X)-\phi)e^{-Q_p(X)/\theta}\;,\nonumber\\
\end{eqnarray}
which we may write as
\begin{equation}
\label{eq3459}
Z(\theta,A_w,A_p)=\frac{1}{A_p\ A}\ e^{-Q_F(\theta,A_w,A_p)/\theta}\;.
\end{equation} 

We may repeat this procedure one more time.  We rewrite the partition 
function $Z(\theta,A_w,A_p)$, equation (\ref{eq3458}) as
\begin{equation}
\label{eq34600}
Z(\theta,A_w,A_p)=\int_{-\infty}^{+\infty}\ dQ_p e^{-Q_p/\theta}\ Z(Q_p,A_w,A_p)\;,
\end{equation}
where 
\begin{eqnarray}
\label{eq3460}
&&Z(Q_p,A_w,A_p)=\int\ dX\ \nonumber\\
&&\delta(Q_p(X)-Q_p)\ \delta(A_w(X)-A_w)\ \delta(A_p(X)-A_p)\;.\nonumber\\
\end{eqnarray}
This {\it microcanonical\/} partition function (as $Q_p$ is the control variable) 
is also the (unnormalized) {\it density of states\/}
with respect to the variables $Q_p$, $A_w$ and $A_p$.  
It demonstrates that all states $X$ with the same $Q_p$, $A_w$ and $A_p$ are
equally probable. This brings us back to our starting point: the entropy has its maximum
when all states that comply with the constraints (\ref{eq2600}), (\ref{eq2700}) and (\ref{eq2750}) 
are equally probable.       
 
\subsection{Co-moving velocity}
\label{co-moving}

The co-moving velocity, which is defined in equations (\ref{eq2000}) and (\ref{eq2100}),
constitutes the bridge between the seepage velocities, equations (\ref{eq500}) and (\ref{eq600}),
and the thermodynamics velocities, equations (\ref{eq1300}) and (\ref{eq1400}). By combining
the defining equations for the co-moving velocities with the two equations ensuing from the
Euler scaling assumption for the volumetric flow rate, we express the seepage velocity of
the two fluid species in terms of the average seepage velocity and the co-moving velocity in
equations (\ref{eq2200}) and (\ref{eq2300}).  

Combining equation (\ref{eq2300}) with equation (\ref{eq5100}) based on statistical mechanics, we find
a consistent structure
\begin{equation}
\label{eq5200}
v_n(\sigma,\mu)=v_p\left(\sigma,S_w(\sigma,\mu)\right)-S_w(\sigma,\mu)\left(\mu-v_m(\sigma,\mu)\right)\;,\\
\end{equation}
when assuming that $v_n=v_n(\sigma,\mu)$ and $v_m=v_m(\sigma,\mu)$. Thus, we have expressed the {\it physical\/}
seepage velocity for the non-wetting fluid within the pseudo-thermodynamic formalism we are developing.  The 
corresponding physical seepage velocity for the wetting fluid may then be found e.g.\ by using equation (\ref{eq900}).

The phenomenological form found by Roy et al.\ \cite{rpsh22}, equation (\ref{eq2400}), is consistent
with the assumption for $v_m$, as equation (\ref{eq2400}) then takes the form  
\begin{equation}
\label{eq5300}
v_m(\sigma,\mu)=A(\sigma)+B(\sigma)\mu\;,
\end{equation}
with
\begin{equation}
\label{eq5400}
B(\sigma)=\left(\frac{\partial v_m(\sigma,\mu)}{\partial \mu}\right)_\sigma\;.
\end{equation}
We note that the dependence of $A$ and $B$ on the capillary number is consistent with the two 
coefficients depending on the cluster entropy density $\sigma$. 

Why $v_m$ is linear in $\mu$ is not known.  

\subsection{Maxwell relations}
\label{maxwell}

We now see that the Euler homogeneity approach of Hansen et al.\ \cite{hsbkgv18} was just the tip of an iceberg.
Having anchored the approach in a statistical mechanics, we now have access to a rich formalism that parallels
thermodynamics.

For example, we find the equivalents of the Maxwell relations in ordinary thermodynamics.  We derive just one
in the following.  

We have that
\begin{equation}
\label{eq5500}
\left(\frac{\partial Q_F(\theta,A_w,A_p)}{\partial \theta}\right)_{A_w,A_p}=\Sigma\;.
\end{equation}
We combine this equation with equation (\ref{eq4300}), taking the cross derivatives, and finding
\begin{equation}
\label{eq5600}
\left(\frac{\partial \Sigma}{\partial A_w}\right)_{\theta,A_p}=\left(\frac{\partial \mu}{\partial \theta}\right)_{A_w,A_p}\;,
\end{equation}
which may be written as
\begin{equation}
\label{eq5700}
\left(\frac{\partial \sigma}{\partial S_w}\right)_{\theta}=\left(\frac{\partial \mu}{\partial \theta}\right)_{S_w}\;.
\end{equation}

\section{Fluctuations and agiture}
\label{fluctuations}

Our starting point are equations (\ref{eq3400}) and (\ref{eq3800}), i.e.,
\begin{eqnarray}
\label{eq5800}
&&Q_G(\theta,\mu,\pi)=-\theta\ln[Z(\theta,\mu,\pi)]\nonumber\\
&=&-\theta\ln\left[\int dX e^{-Q_p(X)/\theta+\mu A_w(X)/\theta+\pi A_p(X)/\theta}\right]\;.\nonumber\\
\end{eqnarray}
This immediately gives
\begin{eqnarray}
\label{eq5900}
&&\left(\frac{\partial Q_G(\theta,\mu,\pi)}{\partial \mu}\right)_{\theta,\pi}\nonumber\\
&=&\frac{1}{Z}\int dX A_w(X) e^{-Q_p(X)/\theta+\mu A_w(X)/\theta+\pi A_p(X)/\theta}\nonumber\\
&=&A_w\;.
\end{eqnarray}
Taking the derivative a second time with respect to $\mu$ gives
\begin{eqnarray}
\label{eq6000}
&-&\theta\left(\frac{\partial^2 Q_G(\theta,\mu,\pi)}{\partial \mu^2}\right)_{\theta,\pi}\nonumber\\
&=&\frac{1}{Z}\int dX A_w^2(X)e^{-Q_p(X)/\theta+\mu A_w(X)/\theta+\pi A_p(X)/\theta}\nonumber\\
&-&\left[\frac{1}{Z}\int dX A_w(X)e^{-Q_p(X)/\theta+\mu A_w(X)/\theta+\pi A_p(X)/\theta}\right]^2\nonumber\\
&=&\langle A_w^2\rangle-A_w^2=\Delta A_w^2\;,
\end{eqnarray}
where we have defined the fluctuations $\Delta A_w^2$.
We now combine equations (\ref{eq100}) and (\ref{eq700}) to find
\begin{equation}
\label{eq6010}
A_w=A\ \frac{A_p}{A}\ \frac{A_w}{A_p}=A\phi S_w\;.
\end{equation}
This allows us to transform equation (\ref{eq6000}) into 
\begin{equation}
\label{eq6100}
\Delta (\phi S_w)^2=\frac{\theta}{A}\left(\frac{\partial (\phi S_w)}{\partial \mu}\right)_{\theta,\pi}\;.
\end{equation}

We see that the agiture $\theta$ seems proportional to the fluctuations
$\Delta (\phi S_w)^2$. However, due to the term $(\partial \phi S_w/\partial \mu)_{\theta,\pi}$, the relation
between them is complex.
We may calculate the porosity fluctuations by taking the partition function $Z(\theta,\mu,\pi)$,
equation (\ref{eq3450}), as starting point. We find
\begin{equation}
\label{eq7000}
\Delta\phi^2=\frac{\theta}{A}\ \left(\frac{\partial\phi}{\partial \pi}\right)_{\theta,\mu}\;,
\end{equation} 
where
\begin{equation}
\label{eq7010}
\Delta\phi^2=\frac{1}{A^2}\left[\langle A_p(X)^2\rangle-A_p^2\right]\;.
\end{equation}
We note that the porosity field $\phi$ is a property of the matrix and not the 
flow.  Hence, equation (\ref{eq7000}) gives a direct link between the agiture $\theta$
and the flow pressure $\pi$,
\begin{equation}
\label{eq7020}
\theta=A\Delta\phi^2\ \left(\frac{\partial\pi}{\partial\phi}\right)_{\theta,\mu}\;.
\end{equation}   

\section{Conditions for steady state in a heterogeneous plug}
\label{pressure}

We consider in the following the conditions for steady state in a heterogeneous plug.   

\begin{figure}[tbp] 
  \centering
  \includegraphics[bb=132 141 757 441,width=7cm,keepaspectratio]{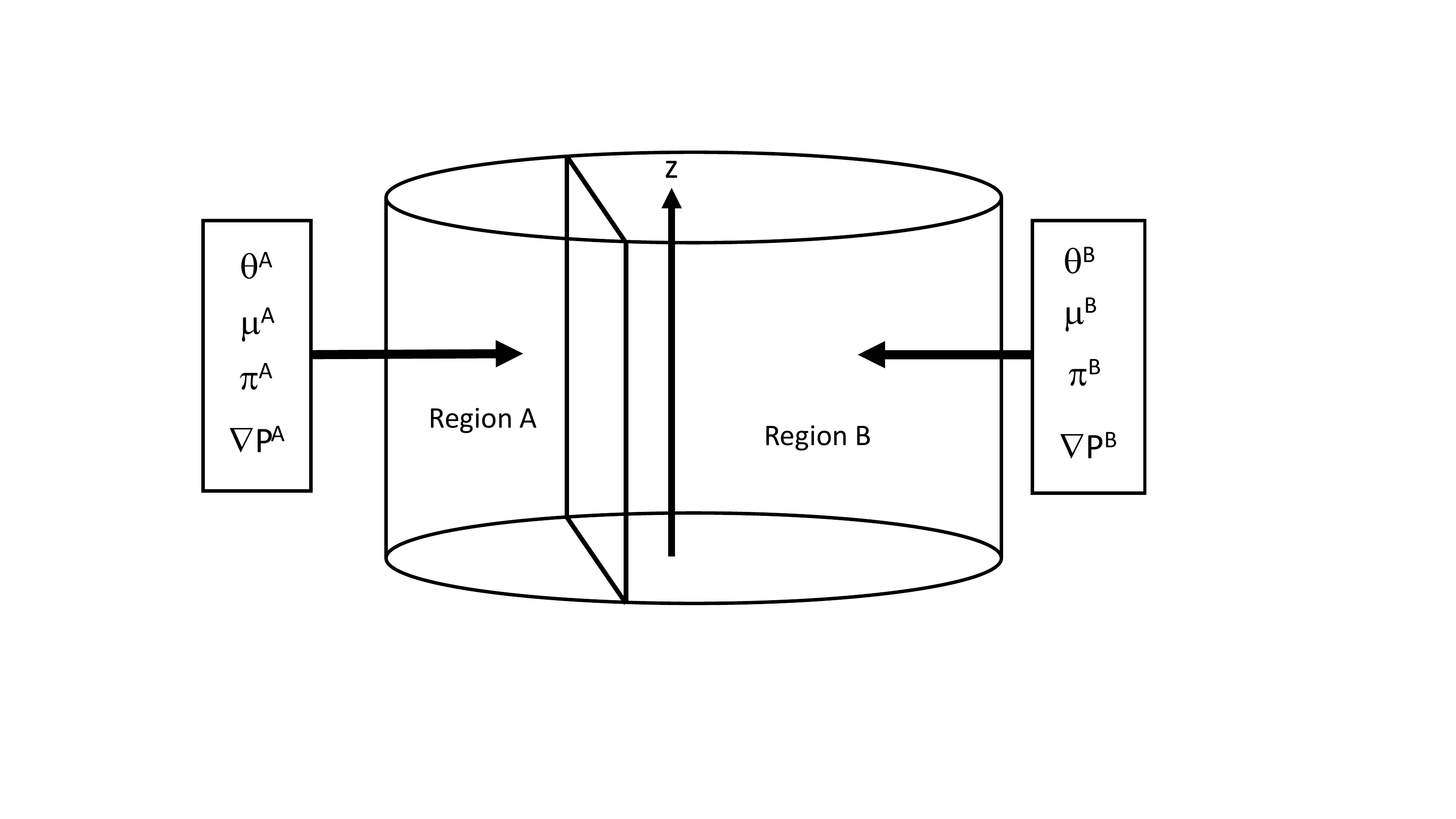}
  \caption{The plug is now divided into a region A and a region B
which differ in composition. The flow in region A is characterized by
the variables $\theta^A$, $\mu^A$, $\pi^A$ and $\nabla P^A$, whereas in 
region B we have  $\theta^B$, $\mu^B$, $\pi^B$ and $\nabla P^B$. 
Steady state flow is attained when $\theta^A=\theta^B$ and $\mu^A=\mu^B$
and $\pi^A=\pi^B$ 
according to equations (\ref{eq100121}), (\ref{eq100123}) and (\ref{eq100123}). 
In addition we have $\nabla P^A=\nabla P^B$.}
  \label{fig4}
\end{figure}

We show in Figure \ref{fig4} a plug that is divided into a region A and a region B,
which have different matrix properties.  The difference may for example be that 
the wetting properties of the matrix in region A differ from those of region B. The full
system, which consists of both regions A and B, is a closed system. 

We will in the following focus on the four quantities $Q_p$, $A_w$, $A_p$ and $\Sigma$ that
describe the flow in the plug. Strictly speaking, we should change our notation, e.g., $Q_p\to{\cal Q}_p$, 
since we are  
considering the entire plug.  However, in order to avoid complicating the notation and therefore making 
the material less accessible, we refrain from making the change.      
   
These four quantities are extensive, i.e., additive.  
This means that we may express them in terms of the two regions A and B.  We have
\begin{eqnarray}
Q_p=Q_p^A+Q_p^B\;,\label{eq10010}\\
\Sigma=\Sigma^A+\Sigma^B\;,\label{eq10011}\\
A_w=A_w^A+A_w^B\;,\label{eq10012}\\
A_p=A_p^A+A_p^B\;.\label{eq10013}\\
\nonumber
\end{eqnarray}

We write $Q_p$ in terms of the control variables $\Sigma$, $A_w$ and $A_p$,
\begin{equation}
\label{eq10014}
Q_p(\Sigma,A_w,A_p)=Q_p^A(\Sigma^A,A_w^A,A_p^A)+Q_p^B(\Sigma^B,A_w^B,A_p^B)\;.
\end{equation}
Since the total volumetric flow rate is conserved in the plug, we must have that 
fluctuations in it must be zero, i.e,
\begin{equation}
\label{eq10015}
\delta Q_p=\delta Q_p^A+\delta Q_p^B=0\;.
\end{equation}

The fluctuations in $Q_p^A$ and $Q_p^B$ come from fluctuations in $\Sigma^A$ and $\Sigma^B$, $A_w^A$ and
$A_w^B$, and $A_p^A$ and $A_p^B$. We express $\delta Q_p^A$ and $\delta Q_p^B$ in terms of the fluctuations of these variables,
\begin{eqnarray}
\label{eq20017}
&\delta Q_p^A(\Sigma^A,A_w^A,A_p^A)&\nonumber\\
&=&\left(\frac{\partial Q_p^A(\Sigma^A,A_w^A,A_p^A)}{\partial \Sigma^A}\right)_{A_w^A,A_p^A}\ \delta\Sigma^A\nonumber\\
&+&\left(\frac{\partial Q_p^A(\Sigma^A,A_w^A,A_p^A)}{\partial A_w^A}\right)_{\Sigma^A,A_p^A}\ \delta A_w^A\nonumber\\
&+&\left(\frac{\partial Q_p^A(\Sigma^A,A_w^A,A_p^A)}{\partial A_p^A}\right)_{\Sigma^A,A_w^A}\ \delta A_p^A\;.\nonumber\\
\end{eqnarray}
Likewise, we have 
\begin{eqnarray}
\label{eq20018}
&\delta Q_p^B(\Sigma^B,A_w^B,A_p^B)&\nonumber\\
&=&\left(\frac{\partial Q_p^B(\Sigma^B,A_w^B,A_p^B)}{\partial \Sigma^B}\right)_{A_w^B,A_p^B}\ \delta \Sigma^B\nonumber\\
&+&\left(\frac{\partial Q_p^B(\Sigma^B,A_w^B,A_p^B)}{\partial A_w^B}\right)_{\Sigma^B,A_p^B}\ \delta A_w^B\nonumber\\
&+&\left(\frac{\partial Q_p^B(\Sigma^B,A_w^B,A_p^B)}{\partial A_p^B}\right)_{\Sigma^B,A_w^B}\ \delta A_p^B\;.\nonumber\\
\end{eqnarray}
 
There is no production of cluster entropy as this entropy is at a maximum.  However, cluster entropy may move between
regions A and B.  Hence, we have
\begin{equation}
\label{eq10016}
\delta \Sigma=\delta \Sigma^A+\delta \Sigma^B=0\;.
\end{equation}

Furthermore, we keep the control variables $A_w$ and $A_p$ fixed.  This is of course not possible experimentally.  
However, as a theoretical concept, it is permissible. Hence, we have that     
\begin{eqnarray}
\delta A_w=\delta A_w^A+\delta A_w^B=0\;,\label{eq10019}\\
\delta A_p=\delta A_p^A+\delta A_p^B=0\;.\label{eq10020}\\
\nonumber
\end{eqnarray}

We now combine equations (\ref{eq10015}) --- (\ref{eq10020}) to find
\begin{eqnarray}
\label{eq20019}
&&\delta Q_p(\Sigma,A_w,A_p)\nonumber\\
&=&\left[\left(\frac{\partial Q_p^A}{\partial \Sigma^A}\right)_{A_w^A,A_p^A}
-\left(\frac{\partial Q_p^B}{\partial \Sigma^B}\right)_{A_w^B,A_p^B}\right] \delta \Sigma^A\nonumber\\
&+&\left[\left(\frac{\partial Q_p^A}{\partial A_w^A}\right)_{\Sigma^A,A_p^A}
-\left(\frac{\partial Q_p^B}{\partial A_w^B}\right)_{\Sigma^B,A_p^B}\right] \delta A_w^A\nonumber\\
&+&\left[\left(\frac{\partial Q_p^A}{\partial A_p^A}\right)_{\Sigma^A,A_w^A}
-\left(\frac{\partial Q_p^B}{\partial A_p^B}\right)_{\Sigma^B,A_w^B}\right] \delta A_p^A=0\;.\nonumber\\
\end{eqnarray}

We now {\it assume that the cluster entropy fluctuations $\delta\Sigma^A$
are independent of the fluctuations in $\delta A_w^A$ and $\delta A_p^A$.\/}  
This leads to 
\begin{eqnarray}
\label{eq100121}
\theta^A&=&\left(\frac{\partial Q_p^A(\Sigma^A,A_w^A,A_p^A)}{\partial \Sigma^A}\right)_{A_w^A,A_p^A}\nonumber\\
&=&\left(\frac{\partial Q_p^B(\Sigma^B,A_w^B,A_p^B)}{\partial \Sigma^B}\right)_{A_w^B,A_p^B}=\theta^B\;,\nonumber\\
\end{eqnarray}
where we have used equations (\ref{eq4000}) and (\ref{eq4150}).

If we now furthermore  {\it assume that the fluctuations in $\delta A_w^A$ and $\delta A_p^A$ are uncorrelated,\/}
we find
\begin{eqnarray}
\label{eq100122}
\mu^A&=&\left(\frac{\partial Q_p^A(\Sigma^A,A_w^A,A_p^A)}{\partial A_w^A}\right)_{\Sigma^A,A_p^A}\nonumber\\
&=&\left(\frac{\partial Q_p^B(\Sigma^B,A_w^B,A_p^B)}{\partial A_w^B}\right)_{\Sigma^B,A_p^B}=\mu^B\;,\nonumber\\
\end{eqnarray}
and
\begin{eqnarray}
\label{eq100123}
\pi^A&=&\left(\frac{\partial Q_p^A(\Sigma^A,A_w^A,A_p^A)}{\partial A_p^A}\right)_{\Sigma^A,A_w^A}\nonumber\\
&=&\left(\frac{\partial Q_p^B(\Sigma^B,A_w^B,A_p^B)}{\partial A_p^B}\right)_{\Sigma^B,A_w^B}=\pi^B\;.\nonumber\\
\end{eqnarray}

These three criteria for steady state flow, equations (\ref{eq100121}), (\ref{eq100122}) and (\ref{eq100123}), 
are analogous to the criteria for two open systems in thermal contact and in equilibrium: here the temperature, pressure 
and the chemical potential must be the same.

It is important to point out the following.  In ordinary thermodynamics, the rule is that the conjugate of 
a conserved quantity is constant in a heterogeneous system at equilibrium.  This is e.g., the argument for the 
temperature being the same everywhere in the system at equilibrium as the entropy is at a maximum.  This is the
same argument as we have used here which leads to equation (\ref{eq100121}).  However, consider two magnets 
with different magnetic susceptibilities in contact which is placed in a uniform magnetic field $H$.  The conjugate of 
the magnetization $M$, which is the magnetic flux $B$, is not equal in the two magnets in contact, $B^A\neq B^B$. 
What goes wrong here is that it is not possible to let 
magnetization fluctuate between the two magnets so that its sum is constant, i.e., $\delta M^A=-\delta M^B$ 
while simultaneously keeping the total internal energy and the total entropy constant, i.e., $\delta U=\delta U^A+\delta U^B=0$ and 
$\delta S=\delta S^A+\delta S^B=0$. 
In our system --- steady-state two-phase flow in a porous medium --- the following question then becomes
central: is it possible to fulfill $\delta Q_p=0$ and $\delta\Sigma=0$, and at the same time have $\delta A_w^A=-\delta A_w^B$ and
$\delta A_p^A=-\delta A_p^B$?  The answer to this question is not obvious.  Only experiments or computations on
models may provide an answer.  If the answer is no, only equation (\ref{eq100121}) will be valid, and not 
equations (\ref{eq100122}) and (\ref{eq100123}).

What about the pressure gradient? We must have that 
\begin{equation}
\label{eq8003}
\nabla P^A=\nabla P^B
\end{equation}  
for steady-state conditions to apply.  If the two pressure gradients were different,
a pressure gradient orthogonal to the flow direction would develop, which would then
divert the flow from the $z$ direction, breaking the assumption of steady-state flow. 

Equation (\ref{eq8003}) is fulfilled if we have that
\begin{equation}
\label{eq8004}
|\nabla P|=|\nabla P|(\theta,\mu,\pi)\;,
\end{equation}
due to  (\ref{eq100121}), and if (\ref{eq100122}) and (\ref{eq100123}) are valid. 

\section{Discussion and Conclusion}
\label{conclusion}

A central problem in the physics of porous media is how to scale up knowledge of the physics at the pore level
to the continuum scale, often referred to as the Darcy scale, where the pores are negligible in size.
The notion of up-scaling is of course only meaningful if we know which coarse-scale physics we are up-scaling to,
and the traditional relative permeability theory has well known weak points. The theory of Hansen et al.\ \cite{hsbkgv18}
is a radically different approach to the coarse-scale physics. This theory is formally similar
to some parts of thermodynamics, but it is the flow rates, and the relations between
flow rates, that are the central players. In its core the theory is based on the volumetric flow rate being an Euler homogeneous
function.

In the present paper, we have developed a statistical mechanics framework for immiscible incompressible two-phase flow.
In this framework, total flow rate plays the same role as energy in ordinary statistical mechanics,
and local steady state flow plays the role of local thermodynamic equilibrium.
We have shown that the pseudo-thermodynamics that follows from the statistical mechanics framework extends the earlier
theory of  Hansen et al.\ \cite{hsbkgv18}, rendering it a complete pseudo-thermodynamics
in the spirit of Edwards and Oakeshot's pseudo-thermodynamics for powders \cite{eo89}.

In the same way that ordinary statistical mechanics serves as a tool for calculating macro-scale
properties from known molecular scale physics, the present statistical mechanics links
the pore scale hydrodynamic description to a continuum scale physics, thus solving the up-scaling problem.
The key link is the partition function $Z(\theta,\mu,\pi)$ defined in equation (\ref{eq3400}).  The integral is over all
\textit{physical fluid configurations,\/} and he physics sits in the measure $dX$ over configurations and
pore structure.

The up-scaling from the intermittent flow of fluid clusters through pore space
to a description with a small number of continuum scale variables admits an immense level of ignorance.
This ignorance is reflected in the pseudo-thermodynamics as the cluster entropy.
We call the conjugate variable to this cluster entropy the agiture.
The agiture measure the level of agitation in the flow. Typically a high agiture
is associated with high flow rates, in the same way high temperature is
associated with high energies in thermal systems.

We have not proposed here any technique as to how the agiture $\theta$ may be measured, 
or controlled, experimentally or computationally. The flow pressure $\pi$ seems also 
difficult to measure. 
Measuring the flow derivative is, on the other hand, seemingly easier to measure. 
In terms of the average seepage velocity and the saturation, it may be written
\begin{equation}
  \mu=\left(\frac{\partial v_p(\sigma,S_w)}{\partial S_w}\right)_\sigma\;.
  \label{eq6600}
\end{equation}
Since we are assuming steady-state flow, the cluster entropy is at a maximum and therefore
constant.  

We also note that we have used five different ensembles in this work, where control parameters
are respectively
\begin{eqnarray}
&&(\theta,\mu,\pi)\;,\nonumber\\
&&(\theta,\mu,A_p)\;,\nonumber\\
&&(\Sigma,\mu,A_p)\;,\nonumber\\
&&(\theta,A_w,A_p)\;,\nonumber\\
&&(\Sigma,A_w,A_p)\;.\nonumber
\end{eqnarray}
The first of these ensembles is easily realizable experimentally as this describes an REA communicating
freely with the rest of the porous medium.  The second one, with $A_p$ controlled, is feasible e.g.\ by
using 3D printing techniques to construct the porous medium.  The two last three ensembles must be seen as
theoretical constructs.  We note, however, that this is standard procedure in thermodynamics in that one
always starts with the Gibbs relation, which relates variations in the internal energy to variations in
the {\it extensive variables\/} irrespective of whether this ensemble is accessible experimentally or not.

We see that a steady-state condition in a heterogeneous porous medium such as that shown in Figure
\ref{fig4} is that the agiture is constant throughout the entire system, equation (\ref{eq100121}).
On the other hand, the two additional equilibrium conditions, equations (\ref{eq100122}) and (\ref{eq100123}),
hinge on additional assumptions that need to be verified experimentally or computationally. In light of the above discussion,
verifying the validity of equation (\ref{eq100122}) is the easier one of the three equations.     

The probability to find a flow configuration $p(X)$ in the REA which is the central to the 
theory we present here.  It was emphasized at the beginning of Section \ref{jaynes} that 
the fluid configuration $X$ refers to the configuration in the REA, not the entire
plane cutting through the plug, ${\cal X}=X\cup X_r$. It is crucial for the theory that $p(X)$ 
does \textit{not\/} depend on the properties of the entire plug.  Recent numerical work by Fyhn et al.\
\cite{fsh22} indicates that this is indeed correct.   

Given all these caveats and open question, the fact remains that the framework we have presented
here, opens up for viewing immiscible two-phase flow in porous media in a different way than before.
We have here divided the
problem into 1.\ constructing a framework by which the concepts that are necessary are put in place,
and 2.\ connecting these concepts to the underlying fluid dynamics and
thermodynamics. This paper accomplishes the first goal.
\section*{Acknowledgment}

The authors thank Dick Bedeaux, Carl Fredrik Berg, Daan Frenkel, Hursanay Fyhn, Signe Kjelstrup, 
Marcel Moura, Knut J{\o}rgen M{\aa}l{\o}y, H{\aa}kon Pedersen, Subhadeep Roy, 
Ole Tors{\ae}ter and {\O}ivind Wilhelmsen for interesting discussions.  
This work was partly supported by the Research Council 
of Norway through its Center of Excellence funding scheme, project number 262644.  

\end{document}